\documentclass[twocolumn]{aastex631}

\usepackage{graphicx}	
\usepackage{amsmath}	
\usepackage{subfigure}
\usepackage{natbib}
\usepackage{color}
\usepackage{tabularx}
\usepackage{longtable}
\usepackage{bm}
\usepackage{threeparttable}
\usepackage{enumerate}
\usepackage[normalem]{ulem}
\usepackage{verbatim}
\usepackage[T1]{fontenc}
\usepackage{soul}

\newcommand{\code}[1]{{\texttt{#1}}}
\newcommand{\tess}{{\it TESS}}
\newcommand{\kepler}{{\it Kepler}}

\newcommand{\gaia}{{\it Gaia}}
\newcommand{\hip}{{\it Hipparcos}}

\begin{document}

\title{Relative Occurrence Rate Between Hot and Cold Jupiters as an Indicator to Probe Planet Migration}

\correspondingauthor{Tianjun Gan}
\email{tianjungan@gmail.com}

\author[0000-0002-4503-9705]{Tianjun~Gan}
\affil{Department of Astronomy, Tsinghua University, Beijing 100084, People's Republic of China}

\author[0000-0001-6870-3114]{Kangrou~Guo}
\affil{Tsung-Dao Lee Institute, Shanghai Jiao Tong University, 520 Shengrong Road, Shanghai 201210, People's Republic of China}

\author[0000-0001-5830-3619]{Beibei~Liu}
\affil{Institute for Astronomy, School of Physics, Zhejiang University, Hangzhou 310027, People's Republic of China}
\affil{Center for Cosmology and Computational Astrophysics, Institute for Advanced Study in Physics, Zhejiang University, Hangzhou 310027, People's Republic of China}

\author[0000-0002-6937-9034]{Sharon X. Wang}
\affil{Department of Astronomy, Tsinghua University, Beijing 100084, People's Republic of China}

\author[0000-0001-8317-2788]{Shude Mao}
\affil{Department of Astronomy, Tsinghua University, Beijing 100084, People's Republic of China}

\author[0000-0003-0426-6634]{Johannes Buchner}
\affil{Max-Planck-Institut für extraterrestrische Physik, Giessenbachstrasse 1, D-85748 Garching, Germany}

\author[0000-0003-3504-5316]{Benjamin J. Fulton}
\affil{Cahill Center for Astronomy \& Astrophysics, California Institute of Technology, Pasadena, CA 91125, USA}
\affil{IPAC-NASA Exoplanet Science Institute, Pasadena, CA 91125, USA}




\begin{abstract}





We propose a second-order statistic parameter $\varepsilon$, the relative occurrence rate between hot and cold Jupiters ($\varepsilon=\eta_{\rm HJ}/\eta_{\rm CJ}$), to probe the migration of gas giants. Since the planet occurrence rate is the combined outcome of the formation and migration processes, a joint analysis of hot and cold Jupiter frequency may shed light on the dynamical evolution of giant planet systems. We first investigate the behavior of $\varepsilon$ as the stellar mass changes observationally. Based on the occurrence rate measurements of hot Jupiters ($\eta_{\rm HJ}$) from the TESS survey and cold Jupiters ($\eta_{\rm CJ}$) from the CLS survey, we find a tentative trend (97\% confidence) that $\varepsilon$ drops when the stellar mass rises from $0.8$ to $1.4\ M_\odot$, which can be explained by different giant planet growth and disk migration timescales around different stars. We carry out planetesimal and pebble accretion simulations, both of which could reproduce the results of $\eta_{\rm HJ}$, $\eta_{\rm CJ}$ and $\varepsilon$. Our findings indicate that the classical core accretion + disk migration model can explain the observed decreasing trend of $\varepsilon$. We propose two ways to increase the significance of the trend and verify the anti-correlation. Future works are required to better constrain $\varepsilon$, especially for M dwarfs and for more massive stars. 

\end{abstract}




\section{Introduction} \label{sec:intro}

About three decades after the first discovery of a hot Jupiter around a Solar-type star \citep{Mayor1995}, the predominant mechanism of their formation remains unclear. Three leading hypotheses have been proposed including in-situ formation, disk as well as tidal migration \citep{Dawson2018}. Thanks to a series of successful ground \citep[e.g,][]{Bakos2004,Pollacco2006,Pepper2007,Chazelas2012} and space \citep[e.g.,][]{Baglin2006,Borucki2010,Howell2014,Ricker2015} surveys, the increasing sample of giant planets enables numerous subsequent statistical studies, which in turn offer unprecedented opportunities to testify these scenarios. 

One of the key demographic parameters is the occurrence rate. Based on the observations from the \kepler\ mission \citep{Borucki2010}, it has been demonstrated that about 0.5\% Sun-like stars host a hot Jupiter with orbital period shorter than 10 days \citep{Fressin2013,Petigura2018,Howard2012}, consistent with the results from early ground-based transit surveys \citep{Gould2006,Kovacs2013}. Despite that radial velocity surveys infer a larger hot Jupiter occurrence rate by a factor of two \citep{Endl2006,Cumming2008,Mayor2011}, such difference might be due to different stellar sample properties like metallicity and binary fraction \citep{Wright2012,Moe2021,Chen2023PAST}. In contrast, cold Jupiters are more common with an occurrence rate about $5-15\%$ \citep{Johnson2010,Wittenmyer2020,Fulton2021,Sabotta2021}, which is predicted by both the core accretion \citep{Pollack1996,Liu2020} and the gravitational instability \citep{Boss1997} planet formation theory. 

Plenty of efforts have been made so far to study the giant planet occurrence rate as a function of stellar type. By splitting the full mixed stellar sample into different mass bins, both \cite{Zhou2019} and \cite{Beleznay2022} claimed a possible anti-correlation between 
the hot Jupiter occurrence rate and host mass. \kepler\ was designed to measure the occurrence rates of different planet systems, which yields good constraint on small planets around M dwarfs \citep{Dressing2013,Dressing2015,Morton2014,Muirhead2015,Mulders2015,Gaidos2016,Hardegree-Ullman2019,Bryson2020a}. However, it has not been able to provide a similar measurement on giant planets due to the limited number of monitored M dwarfs \citep{Johnson2012}. With the help of the Transiting Exoplanet Survey Satellite \citep[\tess;][]{Ricker2015}, \cite{Gan2023} and \cite{Bryant2023} recently extended the occurrence rate function to M dwarfs, where they found that it falls from G dwarfs to the low-stellar-mass end. Meanwhile, the occurrence of cold Jupiters seems to enhance with increasing host mass \citep{Johnson2010,Fulton2021}. Statistical work from \cite{Wolthoff2022} pointed out that the fraction likely peaks around stars with a mass around $1.7\ M_\odot$ and then drops towards stars of $4\ M_\odot$ (see also \citealt{Reffert2015}), consistent with the conclusion from the population synthesis study carried out by \cite{Johnston2023}. Such a stellar mass-dependent feature of giant planet occurrence rate is expected to be a outcome of different amounts of total dust materials. Generally, there are more abundant solids in the protoplanetary disk of high mass stars  \citep[e.g.,][]{Andrews2013}, favoring giant planet formation. Nevertheless, the short disk lifetime of A stars compared with Sun-like stars \citep{Ribas2015} and planet engulfment \citep{Stephan2018} may lead to a low rate of survived giant planets, which probably explains the parabola-shaped giant planet occurrence rate. 

Although the giant planet occurrence rate is supposed to be an intrinsic characteristic, which is bound up with their formation history, later inward migration presumably also leaves some signatures \citep{Lin1996,Burkert2007}. Since stellar mass plays a role in both formation and migration, investigating the occurrence rate of hot and cold Jupiters separately is probably insufficient to untangle the impact from each process. A joint analysis of these occurrence measurements around stars with different masses may provide clues on the dynamical evolution of giant planets. 

To date, various planet formation simulation codes have been developed to investigate the behaviours of planets at the early stage of their formation \citep{Ida2004I,Mordasini2009,Liu2019,Schoonenberg2019,Guo2021,Burn2021,Pan2023}, which also allow for tracking their migration, enabling us to look into how two processes affect giant planet occurrence rate simultaneously. In this paper, we explore the behavior of relative occurrence rate between hot and cold Jupiters at different stellar mass based on statistics from large surveys. We then carry out planetesimal and pebble-driven simulations to examine whether the standard core accretion + disk migration model can reproduce the observation result. A similar study was also carried out by \cite{Su2024} who investigate the mutual occurrence ratio of giant planets based on RV surveys but using different definitions on hot and cold Jupiters. The paper is organized as follows. In Section \ref{obs}, we describe how we construct and combine the observation sample. Section \ref{sim} presents the details of our simulations. We discuss and conclude our findings in Section \ref{dis}. 

\section{Observation results}\label{obs}

In order to make a reliable comparison between observations and simulations, we follow the definition below throughout the paper. We define hot Jupiters (HJs) as planets with mass $100\ M_\oplus\leq M_p\leq 13.6\ M_J$ and semi-major axis $0.02\leq a\leq0.1$ au while cold Jupiters (CJs) as companions with the same mass cut but locating at $1\leq a\leq5$ au. Since we attempt to compare the giant planet occurrence rate $\eta$, the number of planets $N_p$ per 100 stars, across different semi-major axis and stellar mass, we make use of demographic results from the same technique and homogeneous survey for each (HJs and CJs) sample to minimize possible observation bias. 

\subsection{Occurrence Rate of HJs} \label{occ_HJ}

The results of HJ occurrence rate ($\eta_{\rm HJ}$) around different stars are taken from statistical works based on the \tess\ survey \citep{Ricker2015}. We use the occurrence rates of HJs with $0.8\leq R_{p}\leq 2.5\ R_J$ and $0.9\leq P_{b}\leq 10$ days around AFG stars obtained by \cite{Beleznay2022}: $0.29\pm0.05\%$ for A stars (1.4$-$2.3 $M_\odot$), $0.36\pm0.06\%$ for F stars (1.05$-$1.4 $M_\odot$), and $0.55\pm0.14\%$ for G stars (0.8$-$1.05 $M_\odot$). Regarding HJs ($0.6\ R_J \leq R_{p}\leq 2.0\ R_J$ and $1\leq P_{b}\leq 10$ days) around M dwarfs, we use the frequency reported by \cite{Bryant2023}. The authors divided the full M dwarf sample into sub groups and measured individual $\eta_{\rm HJ}$ of $0.137\pm0.097\%$ (0.088$-$0.26 $M_\odot$), $0.108\pm0.083\%$ (0.26$-$0.42 $M_\odot$), and $0.29\pm0.15\%$ (0.42$-$0.71 $M_\odot$). Table~\ref{occvalue} lists the summary of these results. We emphasize that the HJs are defined in the orbital period and planet radius dimensions among transit surveys, which are different from our definition above based on semi-major axis and planet mass. Therefore, we examine the effect that may be induced by such differences. 

Under the definition of \cite{Beleznay2022}, the semi-major axis ranges of HJs are [0.022, 0.111], [0.019, 0.097] and [0.018, 0.088] au around A, F and G stars, respectively. All of them are roughly consistent with our boundary cut [0.02, 0.1] au within about 0.01 au, hence we neglect the difference here. Meanwhile, according to the planet period and stellar mass cut in \cite{Bryant2023}, the semi-major axis ranges of HJs around M stars especially mid-to-late M dwarfs are smaller: [0.011, 0.051], [0.014, 0.063] and [0.016, 0.075] au for the $0.088-0.26\ M_\odot$, $0.26-0.42\ M_\odot$ and $0.42-0.71\ M_\odot$ mass bin. We thus only focus on the $\eta_{\rm HJ}$ around early-M dwarfs with the semi-major axis range closest to our definition for consistency. 

More importantly, we have to be sure the radii range ([0.8, 2.5] $R_J$) defined for the transit occurrence estimates would be consistent with the mass range ([100 $M_{\oplus}$, 13.6 $M_J$]) defined for Jupiters in this work. Note that the radius boundary of HJs adopted in \cite{Beleznay2022} and \cite{Bryant2023} is different ([0.8, 2.5] and [0.6, 2.0] $R_J$). Since no inflated giant planets around M dwarfs with radius above 2.0 $R_J$ have been found yet and the completeness in the statistics will not change significantly given the large transit depth, corresponding to a high signal-to-noise ratio in transit signal search, we thus consider the change of upper boundary from 2.0 to 2.5 $R_J$ will not affect the final occurrence rate estimate. 


However, the difference in the lower radius boundary is expected to have an effect. Among 15 planet candidates included in \cite{Bryant2023}, a total of three objects have radius between 0.6 and 0.8 $R_J$, one of which orbits an early-M dwarf. To correct the radius effect, we compute the occurrence rate of HJs with radius between 0.8 and 2.5 $R_J$ around early-M dwarfs based on the completeness map\footnote{We use the map with radius between 0.8 and 2.0 $R_J$ to estimate the average completeness.} provided in \cite{Bryant2023} following the same procedure therein, and we derive an occurrence rate of $0.23\pm0.15\%$. This calibrated result is close to the original value $0.29\pm0.15\%$ reported in \cite{Bryant2023}. 

Next, we retrieve a list of hot Jupiters with precise mass measurements around AFG and M dwarfs with $0.8\leq R_{p}\leq 2.5\ R_J$ and $1.0\leq P_{b}\leq 10$ days from NASA Exoplanet Archive \citep{Akeson2013}. We find that we will exclude about 7.2\% and 6.7\% HJs around AFG and M dwarfs if using our planet mass cut $\geq 100\ M_\oplus$ instead to define HJs. This difference cannot be corrected statistically unless we have mass measurements for all planet candidates included in both \cite{Beleznay2022} and \cite{Bryant2023}, which are not available and beyond the scope of this study. Consequently, we regard it as a 10\% uncertainty in $\eta_{\rm HJ}$, a small change in the occurrence rate but well within the error bar, when interpreting our results.

Through the above tests, we consider that both the semi-major axis range and the change from radius to mass cut have little impact on $\eta_{\rm HJ}$, which will not affect our final conclusion. We take all $\eta_{\rm HJ}$ around AFG from \cite{Beleznay2022} and the recalculated $\eta_{\rm HJ}$ measurement around early-M dwarfs based on \cite{Bryant2023} into account in the analysis below.


\subsection{Occurrence Rate of CJs}\label{CJ}

The CJ occurrence rate ($\eta_{\rm CJ}$) around stars with different mass are taken from \cite{Fulton2021} delivered by the California Legacy Survey \citep[CLS;][]{Rosenthal2021}. Through the radial velocity measurements for a selected sample of FGKM stars collected under the CLS program, \cite{Fulton2021} determined the occurrence of giant planets beyond the ice line. To be specific, the authors measured the frequency of cold Jupiters ($100 \leq M_p\leq 6000\ M_{\oplus}$) within 1$-$5 au around stars with different mass. 

Although the semi-major axis range is the same, the mass upper boundary (6000 $M_\oplus$) in \cite{Fulton2021} is higher than our threshold (13.6 $M_J$). However, among the total 177 objects, only one sub-stellar companion HD 168443c locates within the semi-major axis limit 1$-$5 au that we pay attention to. HD 168443c has a mass of $17.76\pm0.35\ M_J$ and a semi-major axis of $2.88\pm0.03$ au orbiting a Sun-like star \citep{Rosenthal2021}. Since the difference on $\eta_{\rm CJ}$ after excluding this single object from the total sample is within the $1\sigma$ uncertainty, we consider that it has a negligible effect on our study. Table~\ref{occvalue} summarizes the giant planet occurrence rate results used in this work.

Given the large uncertainty of $\eta_{\rm CJ}$ on M dwarfs due to the very small sample size, we also investigate results from other studies. \cite{Johnson2010} measured a rate about $3.4^{+2.2}_{-0.9}\%$ of M stars with mass below $0.6\ M_{\odot}$ hosting a gas giant with $M_p > 0.3\ M_J$ within 2.5 AU. \cite{Montet2014} determined that about $6.5\pm3.5 \%$ M dwarfs ($0.10-0.64\ M_\odot$) host giant planets within 20 AU with mass $1\leq M_p\leq 13\ M_J$. Based on the CARMENES RV blind survey, \cite{Sabotta2021} found an occurrence rate upper limit of 7\% for giant giant planets ($100-1000\ M_\oplus$) around M dwarfs with period between 100 and 1000 days. Since the definitions of CJs in these studies and the M dwarf mass ranges are fairly different from this work, we choose not to include these measurements. 


\begin{table*}
    \centering
    {\renewcommand{\arraystretch}{1.05}
    \caption{Summary of literature giant planet occurrence rates used in this work.}
    \begin{tabular}{ccccccc}
        \hline\hline
        \multicolumn{7}{c}{Hot Jupiters$^{[1]}$}\\\hline
        Stellar mass range ($M_{\odot}$)       &$0.088-0.26$ &$0.26-0.42$ &$0.42-0.71$ &$0.8-1.05$ &$1.05-1.4$ &$1.4-2.3$\\\hline
        $\eta_{\rm HJ}$ (\%) &$0.14\pm0.10$ &$0.11\pm0.08$ &$0.29\pm0.15$ &$0.55\pm0.14$ &$0.36\pm0.06$ &$0.29\pm0.05$\\\hline
        Semi-major axis range (AU)$^{[2]}$ &$0.011-0.051$ &$0.014-0.063$ &$0.016-0.075$ &$0.018-0.088$ &$0.019-0.097$ &$0.022-0.111$\\\hline \\\hline\hline
        \multicolumn{7}{c}{Cold Jupiters$^{[3]}$}\\\hline
        Stellar mass range ($M_{\odot}$)       &$0.5-0.7$ &$0.7-0.9$ &$0.9-1.1$ &$1.1-1.3$ &$1.3-1.5$ \\\hline
        $\eta_{\rm CJ}$ (\%) &$0.50^{+2.93}_{-0.50}$ &$6.26^{+2.57}_{-1.40}$ &$16.04^{+3.69}_{-2.34}$ &$18.78^{+5.81}_{-4.28}$ &$16.95^{+17.24}_{-3.82}$ \\\hline
        Semi-major axis range (AU) &$1-5$ &$1-5$ &$1-5$ &$1-5$ &$1-5$ \\\hline
    \label{occvalue}    
    \end{tabular}}
    \begin{tablenotes}
       \item[1]  [1]\ Results taken from \cite{Beleznay2022} and \cite{Bryant2023}.
       \item[2]  [2]\ The boundary of semi-major axis is computed based on the stellar mass and planet orbital period range explored in two works. 
       \item[3]  [3]\ Results taken from \cite{Fulton2021}.
    \end{tablenotes}
\end{table*}

\begin{figure*}[htbp]
\centering
\includegraphics[width=0.99\textwidth]{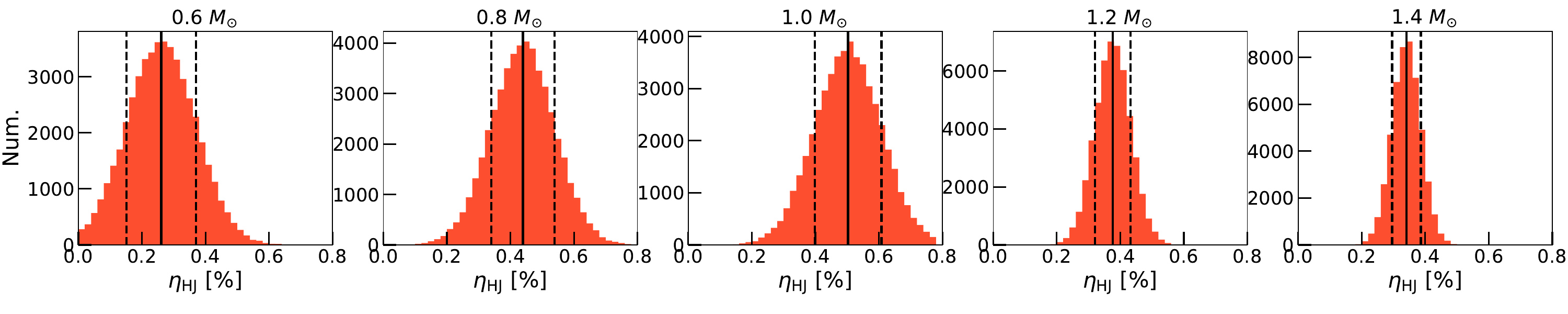}
\caption{The $\eta_{\rm HJ}$ distribution of interpolation results from 50,000 randomly generated data sets (see Section~\ref{relocc} for details). Different panels represent results of different stellar masses. The median and standard deviation of each distribution are shown as the black solid and dashed lines, and listed in Table~\ref{finalresults}.}
\label{hj_intep_plot}
\end{figure*}

\begin{figure*}[htbp]
\centering
\includegraphics[width=0.99\textwidth]{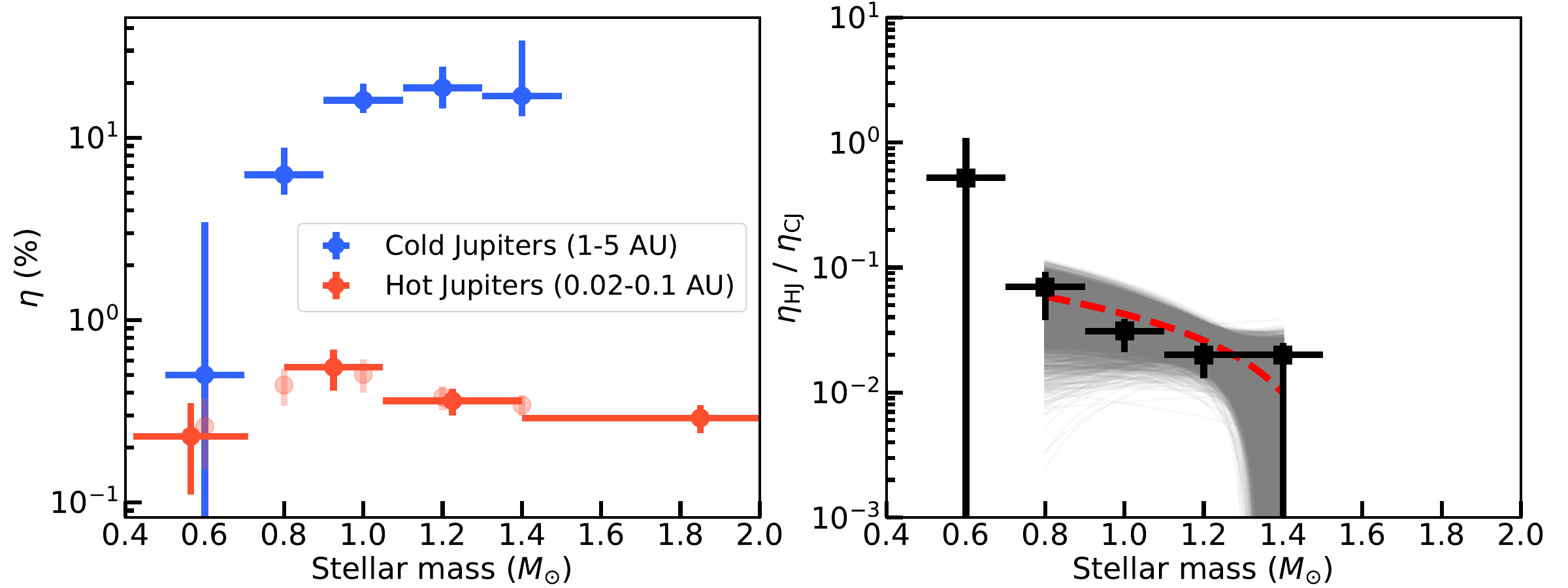}
\caption{{\it Left Panel}: Observed occurrence rate of hot Jupiters ($\eta_{\rm HJ}$) and cold Jupiters ($\eta_{\rm CJ}$) as a function of stellar mass. The results of HJs (red) and CJs (blue) are taken from the \tess\ transit survey \citep{Beleznay2022,Bryant2023} and CLS RV survey \citep{Fulton2021}. Five translucent red dots are the median value of 50,000-time interpolation results (see Section~\ref{relocc}). {\it Right Panel}: Relative occurrence rate $\varepsilon$ between HJs and CJs ($\varepsilon=\eta_{\rm HJ}/\eta_{\rm CJ}$) vs. stellar mass. The gray lines mark the best-fit 1d polynomial functions between $\varepsilon$ and $M_{\star}$ of 10,000 randomly generated data sets while the red dashed line represents the median result. The horizontal uncertainties in both panels mark the stellar mass range. A possible decreasing trend of $\varepsilon$ with increasing stellar mass can be seen.}
\label{occplot}
\end{figure*}

\subsection{Observed Relative Occurrence Rate Between HJs and CJs}\label{relocc}

After two samples are constructed, we investigate the dependence of relative occurrence rate on stellar mass. The relative occurrence rate is defined as the fraction between HJs and CJs ($\varepsilon=\eta_{\rm HJ}/\eta_{\rm CJ}$), which depicts the number ratio (or probability ratio) of giant planets $N_{p, {\rm HJ}}/N_{p, {\rm CJ}}$ within two different semi-major axis ranges around a star. If HJs originally form beyond the snow line and migrate inward, then $\varepsilon$ quantifies the efficiency of such process. A single measurement of $\varepsilon$ in a certain stellar mass bin is less meaningful if the span of two semi-major axis ranges are different (i.e., without normalizing to the same semi-major axis level ${\rm d}N/{\rm d}a$). Nevertheless, through investigating the evolution of $\varepsilon$ with stellar mass, we are able to glimpse into the efficiency of planet formation and migration processes around different types of stars. In principle, the relative giant planet occurrence rate is determined by the planet formation and migration history, both of which will affect the frequency of giant planets at different distance from the hosts.


We first calibrate the occurrence rates of HJs, making the comparison with the occurrence rates CJs is performed in the same stellar mass bins. Assuming each $\eta_{\rm HJ}$ measurement from real observations has a normal distribution, we randomly generate 50,000 sets of $\eta_{\rm HJ}$ of four mass ranges used in \cite{Beleznay2022} and \cite{Bryant2023}, centering at 0.565, 0.925, 1.225 and 1.850 $M_{\odot}$\footnote{These are the central values of each stellar mass bin (See Table~\ref{occvalue}).}. We then linearly interpolate these outputs to five stellar mass bin 0.6, 0.8, 1.0, 1.2 and 1.4 $M_{\odot}$ adopted by \cite{Fulton2021}, which are the central values with a bin size of 0.2 $M_{\odot}$. Figure~\ref{hj_intep_plot} illustrates the $\eta_{\rm HJ}$ distribution of our interpolations. We adopt the median value and standard deviation of each distribution as the occurrence rate and uncertainty used for further analysis. Finally, we compute the $\varepsilon$ of each stellar mass bin by dividing the $\eta_{\rm HJ}$ and $\eta_{\rm CJ}$, where the uncertainty is determined through error propagation. We list $\eta_{\rm HJ}$, $\eta_{\rm CJ}$ and the derived $\varepsilon$ in Table~\ref{finalresults}.

\begin{figure}
\centering
\includegraphics[width=0.49\textwidth]{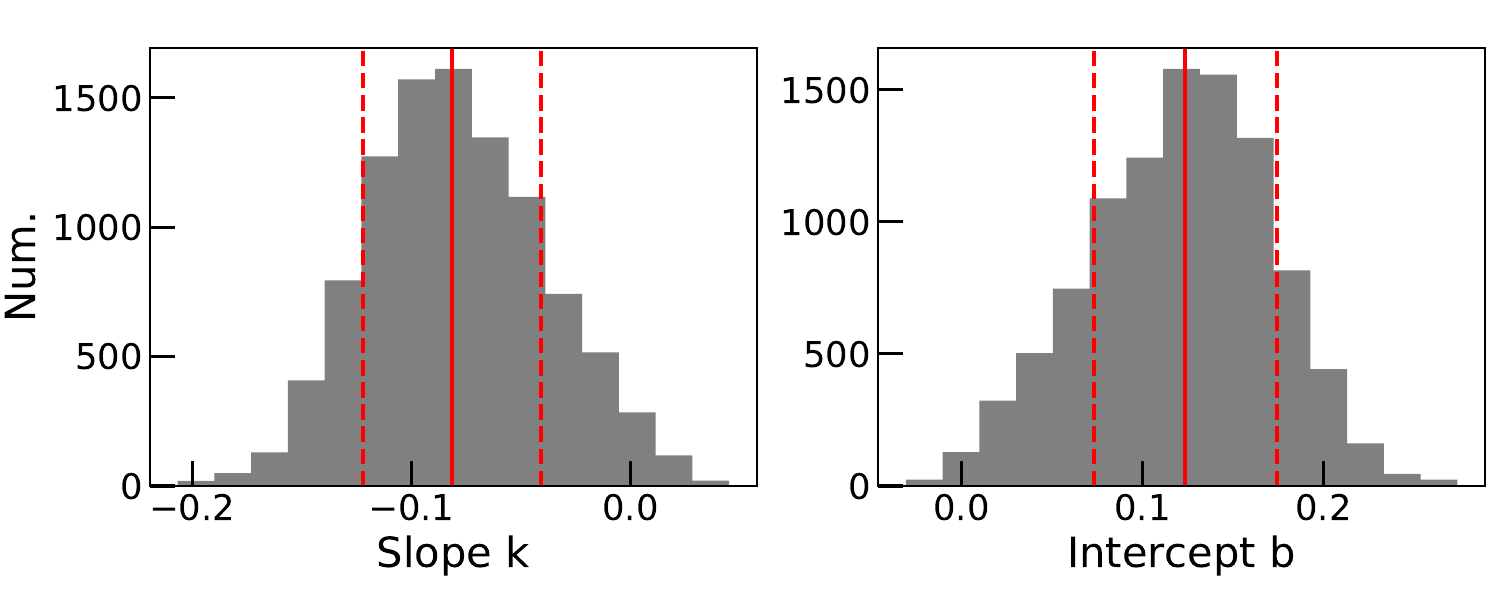}
\includegraphics[width=0.49\textwidth]{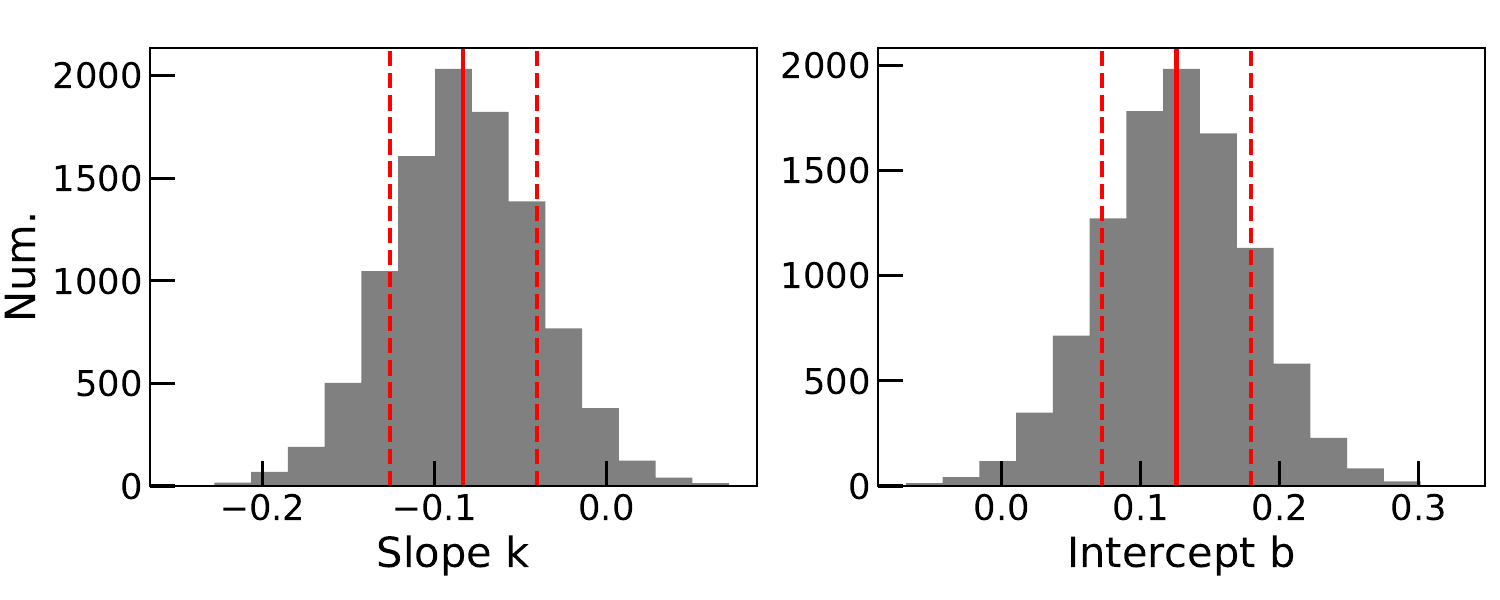}
\caption{The slope and intercept distributions of the linear function fits to 10,000 sets of randomly generated $\varepsilon$ based on observations shown in Figure~\ref{occplot}. The top panel shows the results from simulations based on two half normal distributions while the bottom panel is the outputs from a continuous skewed normal distribution (see Section~\ref{relocc} for details). The red solid and dashed lines represent the median and standard deviation of the distributions. About 97\% data sets have decreasing trends between relative occurrence rate $\varepsilon$ and stellar mass $M_{\star}$.}
\label{slope_plot}
\end{figure}

Figure~\ref{occplot} shows the individual occurrence rate $\eta$ and the relative occurrence $\varepsilon$ as a function of stellar mass. Although only limited measurements are available and the $\varepsilon$ of M stars is poorly constrained due to the large uncertainty of $\eta_{\rm CJ}$, we spot a possible downward trend of $\varepsilon$ with increasing stellar mass (see the right panel of Figure~\ref{occplot}), consistent with the findings reported in \cite{Su2024}. We next examine the significance of the decreasing trend through two ways. Since the result on M dwarfs is highly uncertain, we exclude this data point during our analysis. First, we measure the Pearson’s r correlation coefficient between the other four $\varepsilon$ measurements and stellar mass, where we find a weak anti-correlation ($r=-0.88$) with a p-value of 0.12. We also carry out Monte Carlo simulations based on the derived $\varepsilon$ values listed in Table~\ref{finalresults}. Since the final results of $\varepsilon$ have asymmetrical lower $\sigma_{\varepsilon, {\rm lerr}}$ and upper $\sigma_{\varepsilon, {\rm uerr}}$ uncertainties, we randomly draw data sets of $\varepsilon$ of each stellar mass bin from two half-normal distributions, and select physically meaningful positive $\eta$ values. We fit a linear function between $\varepsilon$ and stellar mass: $\varepsilon=k\cdot (M_{\star}/M_{\odot})+b$. The best-fits are found through the least square method using \code{scipy.optimize}. We repeat the same step for 10,000 times and record the best-fit slope $k$ and intercept $b$ of each simulated data set. We find that 96.7\% of all fitted curves are descending with stellar mass. We also sample from skewed normal distributions using \code{scipy.stats.skewnorm} with 1 sigma quantiles matching the observational error bars for each stellar mass bin, and fit linear functions between $\varepsilon$ and stellar mass. We find similar results that 97.1\% of simulated data sets have negative slopes. For a Gaussian
probability distribution, this probability, considering both tails,
corresponds to a $2.2\sigma$ confidence level. We show the distribution of best-fit slopes we obtained in Figure~\ref{slope_plot}, where the best-fit relation is
\begin{equation}
    \varepsilon=\frac{\eta_{\rm HJ}}{\eta_{\rm CJ}}=[-0.08\pm0.04]\cdot \left(\frac{M_{\star}}{M_{\odot}}\right)+[0.12\pm0.05].
\end{equation}
At this point, we can only report a tentative anti-correlation trend between $\varepsilon$ and stellar mass. We provide two ways to increase the significance of the signal in Section~\ref{future}.

If such a trend indeed exists and is confirmed by future observations, it might be owing to the dependence of the growth as well as migration timescale of giant planets on the stellar mass, which allows testing the dominant roles of theoretical models and hypotheses. Therefore, it motivates the simulations we carried out in Section \ref{sim}.

\begin{table*}
    \centering
    {\renewcommand{\arraystretch}{1.05}
    \caption{Final individual and relation occurrence rates used in this work.}
    \begin{tabular}{cccccc}
        \hline\hline
        Stellar mass bin ($M_{\odot}$)$^{[1]}$       &0.6 &0.8 &1.0 &1.2 &1.4\\\hline
        $\eta_{\rm HJ}^{[2]}$ (\%) &$0.261\pm0.109$ &$0.439\pm0.099$ &$0.503\pm0.105$ &$0.376\pm0.056$ &$0.341\pm0.045$\\\hline
        $\eta_{\rm CJ}^{[3]}$ (\%) &$0.50^{+2.93}_{-0.50}$ &$6.26^{+2.57}_{-1.40}$ &$16.04^{+3.69}_{-2.34}$ &$18.78^{+5.81}_{-4.28}$ &$16.95^{+17.24}_{-3.82}$\\\hline
        $\varepsilon=\eta_{\rm HJ}/\eta_{\rm CJ}$ &$0.522^{+0.565}_{-0.522}$$^{[4]}$ &$0.070^{+0.022}_{-0.032}$ &$0.031^{+0.008}_{-0.010}$ &$0.020^{+0.005}_{-0.007}$ &$0.020^{+0.005}_{-0.020}$\\\hline
    \label{finalresults}    
    \end{tabular}}
    \begin{tablenotes}
       \item[1]  [1]\ These are the central value with a bin size of $0.2\ M_{\odot}$.
       \item[2]  [2]\ All $\eta_{\rm HJ}$ here are the median value from the interpolation results shown in Figure~\ref{hj_intep_plot}.
       \item[3]  [3]\ Same as the results listed in Table \ref{occvalue}, taken from \cite{Fulton2021}.
       \item[4]  [4]\ We exclude this point during our analysis since it is poorly constrained due to the large uncertainty of $\eta_{\rm CJ}$ on M dwarfs reported in \cite{Fulton2021}.
    \end{tablenotes}
\end{table*}

\section{Simulation Results} \label{sim}


\begin{figure*}
    \centering
    \includegraphics[width=0.97\textwidth]{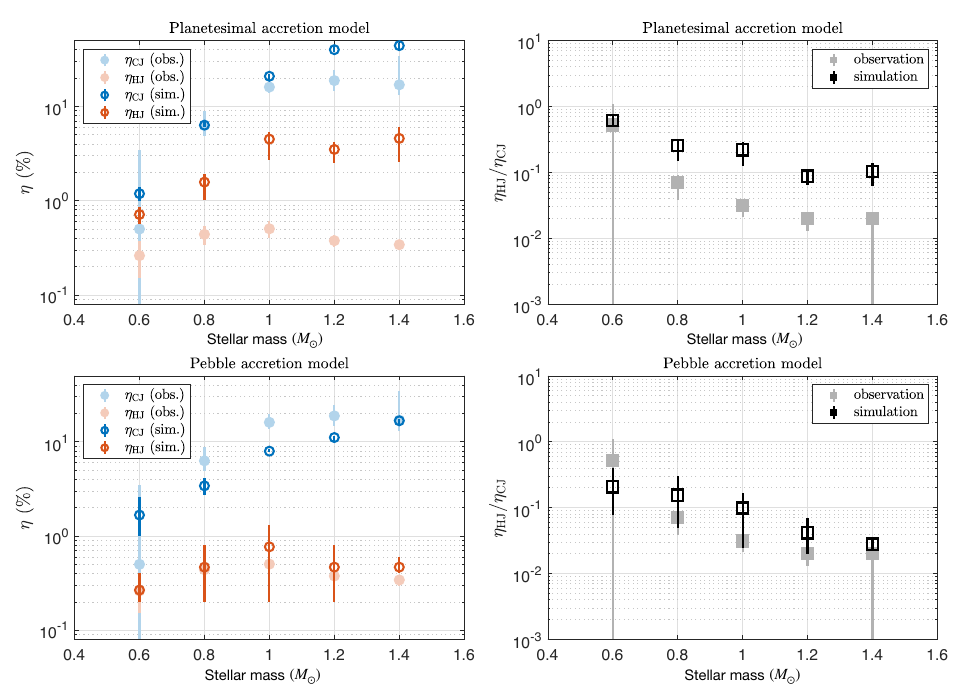}
    \caption{{\it Left Panels}: Simulated occurrence rate of hot Jupiters ($\eta_{\rm HJ}$) and cold Jupiters ($\eta_{\rm CJ}$) as a function of stellar mass. {\it Right Panel}: Relative occurrence rate $\varepsilon$ between HJs and CJs ($\varepsilon=\eta_{\rm HJ}/\eta_{\rm CJ}$) vs. stellar mass. The top and bottom panels show results from the planetesimal and pebble accretion models. The markers show the mean values, and the error bars show the scatter of the results from three sets of simulations. The filled markers represent data from observations as in Fig. \ref{occplot}, and the open markers represent the simulation results.}
    \label{fig:sim}

\end{figure*}


\subsection{Timescale Analysis}

 We first provide an analyitcal estimates on how the growth and disk migration of the planets depend on stellar mass. These two timescales can be expressed as   
 \begin{equation}
  \tau_{\rm mig} =  f_{\rm mig}^{-1} \left( \frac{M_{\star}}{M_{\rm p}} \right) \left( \frac{M_{\star}}{\Sigma_{\rm g} r^{2}} \right) h_{\rm g}^{2} \Omega_{\rm K}^{-1} \propto M_{\star}^{3/4},
  \label{eq:taumig}
\end{equation}
 \begin{equation}
  \tau_{\rm grow} =\frac{ M_{\rm p} }{\dot M_{\rm PA}} = \frac{M_{\rm p}}{\dot M_{\rm peb} \epsilon_{\rm PA}}\propto M_{\star}^{2/3-\beta},
  \label{eq:tauPA}
\end{equation}
where $f_{\rm mig}$ is the migration prefactor, $h_{\rm g}$ is the disk aspect ratio, $\Omega_{\rm K}$ is the Keplerian angular velocity, $\Sigma_{\rm g}$ is the gas density, $r$ is the radial distance, $\dot M_{\rm g}$, $\dot M_{\rm peb} $ are the gas flux and pebble flux  ($\dot M_{\rm peb} {\propto}\dot M_{\rm g}{\propto}M_{\star}^{\beta}$ is assumed), and $\epsilon_{\rm PA}$ is the pebble accretion efficiency ($\epsilon_{\rm PA}{\propto} (M_{\rm p}/M_{\star})^{2/3}$) in the $2$D shear regime \citep{Liu2018}.
In the latter part of the above derivations we only consider the stellar mass scaling and adopt $\Sigma_{\rm g}$ and $h_{\rm g}$ from the inner viscous heating disk region of \cite{Liu2019}.  
Observations indicate that $\beta{\approx}1{-}1.8$ \citep{Hartmann2016}. As such, we can see from Eqs.  \ref{eq:taumig} and \ref{eq:tauPA} that 
the migration timescale increases with $M_{\star}$ while the growth timescale decreases with $M_{\star}$.  In other words, planets grow faster with slow migration speed as their stellar host mass increases\footnote{ Although this derivation is based on pebble accretion, this anti-correlated $M_{\star}{-}\tau_{\rm grow}$ scaling also applies to planetesimal accretion (see Eq.10 of \citealt{Liu2020}). Besides, gas accretion does not alter this $M_{\star}$ scaling. This is due to the Kelvin-Helmholtz contraction following a runaway fashion, which relies on the planet's mass and envelope opacity, without involving any $M_{\star}$ dependence \citep{Ikoma2000}.}. This consequently results in a decreasing trend of $\varepsilon$ with $M_{\star}$.

To further test the above hypothesis, we run planetesimal and pebble accretion population synthesis simulations, examining whether the observed correlation can be reproduced numerically under the core accretion and disk migration framework. In general, the planetesimal accretion model assumes that protoplanets grow by pairwise collisions and accumulation of km-size planetesimals, while the pebble accretion model assumes that protoplanets grow by accreting mm- to cm-size pebbles.
The key difference lies in the role that gas drag plays during the accretion process \citep{Drkazkowska2023}.
In order to make a comparison with observations, for both models we use the same definition of HJ and CJ as stated in Section \ref{obs}, and consider the stellar mass range of 0.6 to 1.4 $M_{\odot}$, in which both $\eta_{\rm{HJ}}$ and $\eta_{\rm{CJ}}$ data are available. We describe the model setups below. 

\subsection{Planetesimal and Pebble Accretion Simulation}
\label{plt}

First, we test whether the planetesimal accretion model can explain the observed correlation between $\varepsilon$ and $M_{\star}$ by performing planet population synthesis simulations using the Ida \& Lin model.
The model details are presented in \citet{Ida_2013} and \citet{Ida_2018}.
The stellar mass has a range of 0.5 to $1.5~M_{\odot}$ and is evenly divided into 5 bins, each with a bin width of $0.2~M_{\odot}$.
Three sets of simulations are performed for each stellar mass bin.
In each bin, the stellar mass is given by a normal distribution with a mean value being the bin center and a dispersion of $0.1~M_{\odot}$.
The retardation factor for orbital migration is $C_1 = 0.05$, which means that the migration speed is reduced by a factor of 20. 
The initial solid and gas surface density resembles the minimum mass solar nebula \citep[MMSN,][]{Hayashi_1981}, with a scaling factor $f$ to control the total disk mass.
The total mass of the protoplanetary disk increases with the stellar mass as $f \propto M_\star^{1.5}$.
The planetesimal disk size ranges from 0.5 to 20 au.  
Following \citet{Ida_2018}, we set the mass of planetesimals and the initial mass of planet embryos to be $10^{20}$ g. The embryos are initially distributed with orbital separations of 10 times the Hill radius of their isolation mass ($r_{\rm{H}} = (m_{\rm{iso}}/3M_\star)^{1/3}a$). The mass accretion rates of the embryos are calculated based on the local disk surface density. For more detailed descriptions of the disk model and their methods, we refer to \citet{Ida_2013} and \citet{Ida_2018}.


We also study the pebble accretion planet population synthesis model by conducting three sets of simulations, each of which contains $2,000$ systems with Monte Carlo sampling of their initial disk and stellar properties. We only consider the growth and migration of a single protoplanet within individual systems. We assume that pebbles have all grown up to a Stokes number of $0.01$, roughly equivalent to a millimeter in size. 
The embryo with $M_{\rm p}=0.01 \ M_{\oplus}$ is placed logarithmically between $0.1$ and $40$ au. Detailed model prescription can be found in \cite{Liu2019}. One notable aspect differed from \cite{Liu2019} is that the type I migration of the planets is adopted from \cite{DAngelo2010}'s $3$D isothermal formula with a retardation factor of $C_{1} {=} 0.1$.
Other initial conditions and distributions of model parameters in the above two models are summarized in Table \ref{simulation_v2}.

Figure \ref{sim} illustrates the simulated $\eta_{\rm HJ}$,  $\eta_{\rm CJ}$ and their ratio as a function of $M_{\star}$.  Both models predict that  $\eta_{\rm HJ}$ and $\eta_{\rm CJ}$ increase with $M_{\star}$. The increasing trend of $\eta_{\rm CJ}$ is steeper than that of  $\eta_{\rm HJ}$, while $\eta_{\rm HJ}$ mildly peaks at solar-mass stars. 
Consequently, the relative occurrence rate decreases with stellar mass, ranging from $0.1{-}1$ around stars of $0.6 \ M_{\odot}$ to $0.02{-}0.1$ around stars of $1.4 \ M_{\odot}$.  These results align well with the observations shown in Figure \ref{obs}. Physically, this implies that a larger fraction of giant planets form around more massive stars, but a small fraction of them undergo disk migration and transition into HJs. 
We note that in both models, one parameter that particularly required tuning in order to match the results with observations is the migration reduction factor $C_1$. 
With larger values of $C_1$, the simulations typically produce a higher fraction of HJs compared with observation due to efficient inward migration. 

\begin{table*}
    \caption{Model parameters in population synthesis models.}
    \begin{tabular}{ccr}
        \hline\hline
        Parameters (Planetesimal model)       &Value &Description      \\\hline
        $\alpha_{\rm{vis}}$, $\alpha_{\rm{acc}}$ & $3\times 10^{-4}$, $3\times 10^{-3}$ &Disk $\alpha$ viscosity (turbulent, accretion). \\
        $k_1$, $k_2$ & 9, 3.5 &Gas contraction$^{[1]}$.\\
        $\rm [Fe/H]$ & $\mathcal{N}^{[2]}$ $(\mu, \sigma^2)$, $\mu=0$, $\sigma=0.2$ & Metallicity.\\
        $\log{(\tau_{\rm{disk}}\ {\rm [yr]})}$ & $\mathcal{N}$ $(\mu, \sigma^2)$, $\mu=6.5$, $\sigma=0.2$ &Disk lifetime.\\
        $\log{(f)}$ & $\mathcal{N}$ $(\mu, \sigma^2)$, $\mu=0$, $\sigma=0.1$ &Disk mass scaling factor.\\
        $p_{\rm{disk}}$ & 1.5 & $ f \propto M_{\star}^{p_{\rm{disk}}}$. \\
        $C_1$ & 0.05 &Migration retardation factor.\\
        \hline
        
        \hline
        \\
        \hline\hline
        Parameters (Pebble model)       &Value &Description     \\\hline
        $\dot M_{\rm g0} \ ( M_\odot  \ \rm yr^{-1})$ & $  10^{{\mathcal{N} (\mu, \sigma^2)}} \times (M_{\star}/M_{\odot})^{1.5}$, $\mu=-7.5$, $\sigma=0.3$ &Disk accretion rate. \\
        $R_{\rm d0} \ \rm (AU)$ & $\mathcal{U}^{[3]}$ (50,200) $\times (M_{\star}/M_{\odot})$ &Disk size. \\
        $\rm [Fe/H]$ & $\mathcal{N}$ $(\mu, \sigma^2)$, $\mu=-0.03$, $\sigma=0.2$ &Metallicity. \\
        $\xi$ & $0.0149 \times 10^{\mathcal{N} (\mu, \sigma^2)}$, $\mu=-0.03$, $\sigma=0.2$ &Pebble-to-gas flux ratio.\\
        $\alpha_{\rm g}$ &  $10^{-2}$  &Global viscous coefficient. \\
        $\alpha_{\rm t}$ &  $10^{-3}$ &Midplane turbulent strength.\\
        $\tau_{\rm s}$  &  $10^{-2}$ &Pebbles' Stokes number.\\
        $C_1$ & 0.1 & Migration retardation factor.\\
        \hline 

        \hline
    \end{tabular}
    \begin{tablenotes}
    \item[1]  
    [1] The gas contraction timescale is $\tau_{\rm{KH}} \simeq 10^{k_1} (M_{\rm{p}}/M_{\oplus})^{-k_2}~\rm{yr}$.
    
    [2] $\mathcal{N}$($\mu\ ,\ \sigma^{2}$) means a normal distribution with mean $\mu$ and standard deviation $\sigma$. 
    
    [3] $\mathcal{U}$(a\ , \ b) stands for a uniform distribution between $a$ and $b$.
    \end{tablenotes}
    \label{simulation_v2}
\end{table*}

\section{Discussions and Conclusions} \label{dis}

In this paper we present a key indicator, the relative occurrence rate between hot and cold Jupiters ($\varepsilon = \eta_{\rm HJ}/\eta_{\rm CJ}$), to trace the formation and evolution of giant planet systems. Combining the statistics based on \tess\ and CLS surveys, we identify a tentative decreasing trend of $\varepsilon$ with increasing stellar mass (97\% confidence). We find that this trend could be explained by different planet growth and migration time scales around different stars through state-of-the-art planetesimal and pebble accretion simulations. Both simulations quantitatively reproduce the observed correlation between stellar mass and $\eta_{\rm HJ}$, $\eta_{\rm CJ}$ as well as $\eta_{\rm HJ}/\eta_{\rm CJ}$.

\subsection{Caveats}

There are several caveats that we do not account for including (1) comparing occurrence rates from two techniques; (2) the statistical methodologies used in three works that this work is based on are different; (3) theoretical models that simplify the retention or destruction of planets near the inner edge of the disk.


A major caveat of this work is that we are comparing $\eta_{\rm CJ}$ from RVs and $\eta_{\rm HJ}$ from transit surveys, and $\eta_{\rm HJ}$ is known to differ by a factor of about two between these two methods. Ideally, one should use $\eta_{\rm CJ}$ and $\eta_{\rm HJ}$ estimates from the same method, but limitations from occurrence rate studies to date prohibit such an attempt. Typical RV studies have a limited sample in HJ detections and thus do not report $\eta_{\rm HJ}$ across stellar types \citep{Wright2012,Fulton2021}, and transit surveys typically do not have the long baseline to recover a large enough sample of CJs \citep{Hsu2019}. 
Worse still, transiting cold Jupiters around M dwarfs have never been detected, making $\eta_{\rm CJ}$ at M dwarfs not available via transits.

To match the $\eta_{\rm HJ}$ from RVs, in principle, one could apply a correction factor to the $\eta_{\rm HJ}$ estimates from transits. The difference between $\eta_{\rm HJ}$ from RV and transits arises from the metallicity \citep{Wright2012}, binary fraction \citep{Moe2021}, age \citep{Chen2023PAST} distributions of stellar samples, or, most likely, a combination of three factors. \cite{Chen2023PAST} found that the discrepancy of $\eta_{\rm HJ}$ between transit and RV surveys could be explained after accounting for the metallicity and age effect (and thus much less from the binarity effect). Such an effect is not yet seen to vary widely across stellar types \citep{Miyazaki2023}, hence it may lead to a similar occurrence rate shift at different stellar masses if we calibrate $\eta_{\rm HJ}$ from transit to RV studies, which will not significantly change the decreasing trend of $\varepsilon$. Although photometric surveys like \kepler\ and \tess\ have large homogeneously selected stellar samples, the stellar catalog of long-term RV surveys might be biased due to human factors and multiple selection criterias, which will lead to risks on further statistics. However, the CLS program selected targets in a statistical sense like a blind survey as California Planet Search \citep[CPS;][]{Howard2010} without bias toward stars more likely to host planets according to metallicity information or previous data \citep[see Section 2 in][]{Rosenthal2021}, making such a situation less serious.

Secondly, the methodologies used in \cite{Fulton2021}, \cite{Beleznay2022} and \cite{Bryant2023} to derive giant planet occurrence rates are different. \cite{Beleznay2022} utilize a Beta distribution model as in \cite{Zhou2019}, where the number of observed planets is computed with a false positive rate according to the dispositions of planet candidates and a weighting parameter of the probability that a star falls in a specific mass bin. The effective number of stars is determined based on the detection sensitivity and completeness through injection and recovery tests. The statistical method applied in \cite{Bryant2023} is similar to \cite{Beleznay2022} but estimating false positive rate through a Bayesian way instead, and without taking the aforementioned weighting parameter into consideration. In terms of cold Jupiters, \cite{Fulton2021} employed the hierarchical Bayesian methodology described in \cite{Hogg2010} and \cite{Foreman-Mackey2014}, which is different from the other two works. We note that we do not adjust these difference in this work and we assume the final occurrence rate outputs ($N_p$ per 100 stars) are not biased between statistical methodologies.

From the simulation side, we only test if the trend can be explained by a single formation mechanism of CJ and HJs: via core accretion plus disk migration. High-eccentricity migration and disk instability are also expected to affect the formation of HJ and CJ systems \citep{Dawson2018}. Theoretical simulations under these frameworks to examine whether they could reproduce the decreasing trend of $\varepsilon$ might be a way to figure out if they played a dominant role compared with disk migration. A major limitation of our simulation is that the models employed are simplified in the sense of describing the evolution of planets after migrating to the inner edge of the protoplanetary disk, because other than the balance between growth and migration as the stellar mass changes, the dependence of the retention efficiency of HJs by the inner disk cavity on the stellar mass may also contribute to the observed decreasing trend of $\varepsilon$. 
For simplicity, the Ida \& Lin model assumes a fixed location for the inner disk edge that does not depend on the central body mass.
In the pebble accretion model, a static inner disk edge is assumed to approximate to the co-rotation radius of the central star.
In both models, the inner edge of the disk serves as the location to halt the inward migration of planets. 
In principle, the inner cavity radius is determined by a balance between the stellar magnetic torque and the gas accretion torque \citep{Koenigl1991,Lin1996,Liu2017}. 
For massive stars ($\gtrsim 1.5~M_{\odot}$), the magnetic fields can be too weak to clear up a magnetospheric cavity in the inner disk due to the lack of convective zones, making it difficult to halt the inward migration of HJs.
In this regard, giant planets are more easily trapped at the inner disk edge around low-mass dwarfs, while they are more likely to migrate inward and eventually collide with the central stars in systems around more massive stars.
In addition, several other mechanisms that could potentially destroy these close-in planets, such as the tidal interaction and disruption \citep{Liu2013} or engulfment of planets \citep{De2023} by the central star, have not been considered in both models.

\subsection{Relative Occurrence Rate Results from \kepler}\label{kepler_results}

The \kepler\ mission has produced lots of occurrence rate results for giant planets around FGK stars \citep{Fressin2013,Howard2012,Petigura2018,Hsu2019,Kunimoto2020}. The completeness as well as reliability of occurrence rates are also well-studied \citep{Burke2015,Bryson2020b}. Therefore, it is worth looking into these results in a similar way and revisiting the trend between the relative occurrence rate $\varepsilon$ and stellar mass $M_\star$. 

\begin{figure}
\centering
\includegraphics[width=0.49\textwidth]{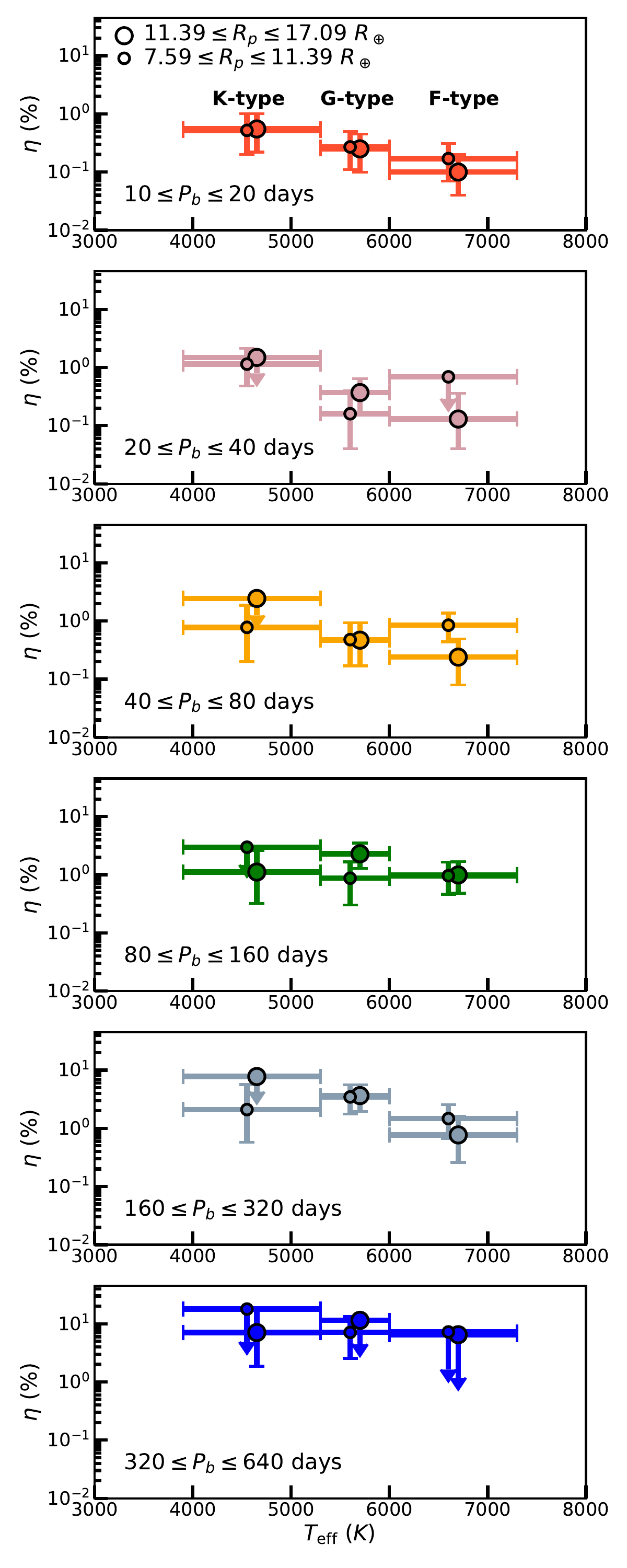}
\caption{Giant planet occurrence rate vs. stellar effective temperature based on \kepler\ mission taken from \citet{Kunimoto2020}. Different panels show results of different period bins. The marker size scales with planet radius ranges shown on the top left in the first panel. We add small shifts to the horizontal coordinates for clarity.}
\label{kepler_result}
\end{figure}

As a reference, we retrieve the results of $\eta$ from \cite{Kunimoto2020}, who presented the planet occurrence rates over the period–radius grid recommended by the Study Analysis Group (SAG) 13 of the NASA Exoplanet Exploration Program
Analysis Group (ExoPAG). The whole stellar sample was divided into three categories based on the suggested $T_{\rm eff}$ limits from \cite{Pecaut2013}: F- ($6000\leq T_{\rm eff}\leq 7300$ K), G- ($5300\leq T_{\rm eff}\leq 6000$ K) and K-type ($3900\leq T_{\rm eff}\leq 5300$ K) stars, different with the selection according to stellar mass used in this work. For the giant planet group, there are $\eta$ measurements in two radius (7.59–11.39 and 11.39–17.09 $R_\oplus$) and six period (10–20, 20–40, 40–80, 80–160, 160–320, 320–640 days) bins (see Figure~\ref{kepler_result}). 

The uncertainties of $\eta$ are higher compared with the results from \tess\ listed in Table~\ref{finalresults}, which will consequently lead to higher uncertainties on $\varepsilon$ if compared with the RV result from CLS. In principle, one can homogeneously measure $\varepsilon$ based on the existing \kepler\ occurrence rate results from the literature, though within two semi-major axis ranges close to the host star. However, we note that the differences of the semi-major axis boundary of three stellar types will be more significant ($\sim 0.4$ AU) towards large period bins when performing the conversion. With these in mind, we compute the $\varepsilon$ based on $\eta$ from the first period bin and others, attempting to investigate the relation between $\varepsilon$ and $T_{\rm eff}$. The results are presented in Figure~\ref{kepler_epsilon_result}. Due to the large uncertainty of $\varepsilon$, we are not able to claim any trend.

\subsection{Future Prospects}\label{future}

At present, the confidence level of the decreasing trend between $\varepsilon$ and stellar mass $M_\star$ is not high. Future observations are required to revisit this parameter, and confirm or rule out the anti-correlation. Here, we suggest two possible ways to improve the significance of the trend.

Due to the limited sample size of CJs around M dwarfs, the constraint on their occurrence rate still has a large uncertainty, especially if attempting to group them into different stellar mass and semi-major axis bins. Including additional $\eta_{\rm CJ}$ data points of M dwarfs with stellar mass below $0.65\ M_\odot$, in particular mid-to-late M stars, would make the behaviour of $\varepsilon$ more clear. However, we emphasize that such measurements should be done under the same definition of CJs ($100\ M_\oplus \leq M_p\leq 13.6\ M_J$, $1\leq a\leq 5$ au) otherwise they cannot be directly compared with the results in this work. Ground-based blind RV surveys, notably in the near-infrared band, definitely provide a pathway to constrain the $\eta_{\rm CJ}$ around M dwarfs \citep[e.g.,][]{Morales2019,Stefansson2023}. Microlensing \citep{Mao1991,Gould1992} and direct imaging \citep{Montet2014,Zhang2021COCONUTS} have also shown their capability in discovering cold giants around M stars, though the detection rates of these two techniques are relatively low. 

Astrometry is a promising route to efficiently find long-period planets in the near future. Based on the absolute astrometry measurements from \hip\ \citep{Perryman1997,vanLeeuwen2007} and \gaia\ \citep{Gaia2018,Gaia2021} as well as ground RVs, a considerable number of CJs have been confirmed with true mass obtained \citep[e.g.,][]{Feng2022,Sozzetti2023}. With the astrometry method, we are able to determine $\eta_{\rm CJ}$ across stars with a wide range of mass, providing a check independent of RV \citep{Perryman2014}. The astronomic signal is proportional to the mass ratio of the planetary system, making it sensitive to giant planets around low-mass stars. Moreover, the detections of planets around more massive stars, for example A-type stars \citep[e.g.,][]{CollierCameron2010,Gaudi2017,Zhou2019}, are limited by their shallow transit depth and small RV amplitude. Such constraints are slightly mitigated when tracking the stellar motions. Extending the $\varepsilon$ function to A and F stars with mass between 1.4 and 2.0 $M_\odot$ will enable the exploration toward the high-mass end. Although the \gaia\ DR4 time-series astrometric data would not be available before the end of 2025, the non-single-star catalog with two body orbit solutions released during DR3 \citep{Gaia2023,Halbwachs2023,Holl2023} have already enabled the true mass determination for several CJs \citep{Winn2022,Gan2023gaia}. 




Finally, we suggest using similar definitions of both HJs and CJs in future demographic works, which will enable joint analysis across different parameter space, making the comparisons between different results easier.

\section{Acknowledgments}

We thank Doug Lin, Shigeru Ida and Yasunori Hori for useful discussions, and two anonymous referees for their comments that improved the quality of this paper.

This work is supported by the National Science Foundation of China (Grant No. 12133005). K. Guo acknowledges the access to Matlab R2021b under the academic license granted by Shanghai Jiao Tong University. BL acknowledges the fundings from the National Natural Science Foundation of China (Nos. 12222303, 12173035, 12147103 and 12111530175), the start-up grant of the Bairen program from Zhejiang University and the Fundamental Research Funds for the Central Universities (2022-KYY-506107-0001,226-2022-00216).

%

\vspace{5mm}
\facilities{\tess, \kepler, Keck}


\software{\code{numpy} \citep{numpy}, \code{matplotlib} \citep{Hunter2007}, \code{scipy} \citep{Virtanen2020}}




\appendix

\section{Relative Occurrence Rate Results from \kepler\ Mission}

We compute the relative occurrence rate $\varepsilon$ based on \kepler-only results from \cite{Kunimoto2020}. We note that the calculations are done in the period bins instead of semi-major axis bins.

\begin{figure}
\centering
\includegraphics[width=0.49\textwidth]{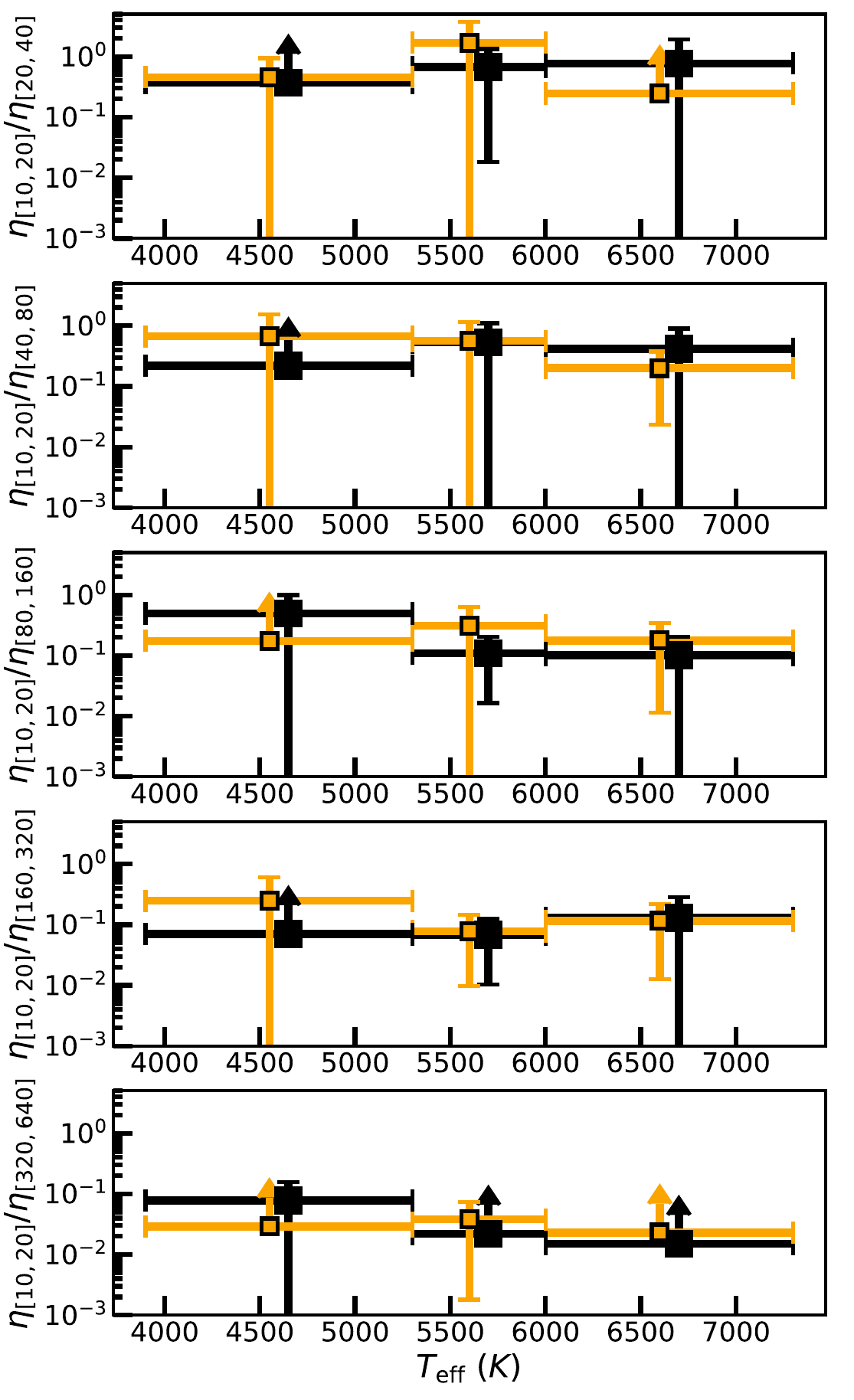}
\caption{Relative occurrence rate based on the occurrence rate results from \citet{Kunimoto2020} as a function of effective temperature. The small orange and large black squares are the results of giant planets within two different radius ranges, $7.59\leq R_{p}\leq 11.39 \ R_{\oplus}$ and $11.39\leq R_{p}\leq 17.09\ R_{\oplus}$  (see Figure~\ref{kepler_result} and Section~\ref{kepler_results} for details). The subscripts of $\eta$ represent the period bins. We add small shifts to the horizontal coordinates for clarity.}
\label{kepler_epsilon_result}
\end{figure}


\bibliography{planet}{}
\bibliographystyle{aasjournal}



\end{document}